\documentclass[oldversion]{aa}
\usepackage{graphicx}
\usepackage{txfonts}

\def\gtabouteq{\,\hbox{\raise 0.5 ex \hbox{$>$}\kern-.77em 
                    \lower 0.5 ex \hbox{$\sim$}$\,$}}       
\def\ltabouteq{\,\hbox{\raise 0.5 ex \hbox{$<$}\kernValues  
from the NASA Extragalactic Database (NED) unless otherwise
indicated.-.77em 
                     \lower 0.5 ex \hbox{$\sim$}$\,$}} 

\begin{document}

\title{PAHs in the Halo of NGC~5529}

\author{J. A. Irwin\inst{1} \and H. Kennedy\inst{1}
	\and T. Parkin\inst{1} \and S. Madden\inst{2}}

\institute{Dept. of Physics, Engineering Physics, \& Astronomy,
 Queen's University, Kingston, Canada, K7L~3N6\\
  \email{irwin@astro.queensu.ca} 
\and 
  CEA/Saclay, Service d'Astrophysique,
Orme des Merisiers, B{\^a}timent 709, 91191 Gif-sur-Yvette cedex, France\\
  \email{smadden@cea.fr}}


\date{Received 00 month 0000; accepted 00 month 0000}

\abstract{
We present sensitive
ISO $\lambda\,6.7~\mu$m observations of the edge-on galaxy,
NGC~5529, finding an extensive
MIR halo around NGC~5529. The emission is dominated by PAHs in
this band.  The PAH halo has an exponential scale height of 3.7 kpc but
can still be detected as far as $\approx\,10$ kpc from the plane 
to the limits of the high dynamic range (1770/1) data.
This is the most extensive
PAH halo yet detected in a normal galaxy.  This halo
shows substructure and the PAHs likely originate from some type of disk
outflow.  PAHs are long-lived in a halo environment and therefore
continuous replenishment from the disk is not required
(unless halo PAHs are also being destroyed or removed), consistent with
the current low SFR of the galaxy.  The PAHs correlate spatially with
halo H$\alpha$ emission, previously observed by Miller \& Veilleux (2003);
both components are likely excited/ionized by in-disk photons that are
leaking into the halo.  The presence of halo gas may be related to the
environment of NGC~5529 which contains at least 17 galaxies in a small
group of which NGC~5529 is the dominant member.  Of these, we have identified
two new companions from the SDSS.}



\maketitle

\section{Introduction}
\label{introduction}

Nearby edge-on galaxies provide an important local laboratory for
understanding physical conditions in the disk-halo
interface.  This critical region separates two  
very different environments over a relatively
small vertical distance.  It plays an important role in a galaxy's
energy balance as well as its chemical evolution, since outflows
from the disk, circulating `fountains', and possibly infalling external
material may each be involved at some point in a galaxy's evolution.  
Moreover, knowledge of the  physical
conditions of outflowing gas in nearby galaxies can
provide important constraints on galaxy formation scenarios,
since outflows (i.e. some form of `feedback') are crucial 
to these models
(e.g. Marri \& White, 2003 and others). Nearby edge-on galaxies have 
now been studied
in many wavebands showing a variety of interstellar medium (ISM)
components above the galactic
disk.  Indeed, in at least one galaxy that has a high star formation
rate (SFR), every component of the 
interstellar medium, including dust, molecular gas,
HI, extra-planar diffuse ionized gas (eDIG, detected by H$\alpha$ 
emission) and X-ray emitting hot gas has been detected above the
 galactic disk (see Lee et al. 2001 and Brar et al. 2003).
In each component, discrete disk-halo vertical structures are seen, 
suggestive of outflow. 

The recent discovery, from 
mid-infrared (MIR) observations, of PAH 
(polycyclic aromatic hydrocarbon) emission\footnote{There is some
uncertainty as to the nature of the carriers of 
the MIR band emission.  In this paper, we adopt
the PAH nomenclature for consistency with other authors
(e.g. Draine \& Li 2007).}
 in the
halo of the low SFR galaxy, NGC~5907
(Irwin \& Madden 2006), has prompted
a search for other galaxies that may show similar emission.  In this
paper, we
present results for the edge-on galaxy, NGC~5529, 
using archival data from the 
Infrared Space Observatory (ISO) in the $\lambda\,6.75$ $\mu$m waveband,
which selects the
$\lambda\,6.2$, $\lambda\,7.7$, and part of the  $\lambda\,8.6$
$\mu$m PAH features (see Fig.~7 of Irwin \& Madden 2006).
These were the only ISO observations taken of this galaxy.
 In Sect.~\ref{ngc5529}, we introduce
the galaxy, Sect.~\ref{observations_reductions} presents
 the observations and data
reduction, Sect.~\ref{results} lists the results, and the discussion
and conclusion are presented in Sects.~\ref{discussion} and
\ref{conclusions}, respectively.

\section{NGC~5529 and its Environment}
\label{ngc5529}

NGC~5529,
 whose basic parameters are listed in 
  Table~\ref{basic_parameters},
is an edge-on galaxy with a prominent dust lane.  It 
 is classified as an Sc galaxy, but
is likely barred (Kregel \& van der Kruit 2004) and a peanut-shaped bulge is
also apparent in the optical image shown in 
Fig.~\ref{n5529_colour}.
At a distance of $D\,=\,43.9$ Mpc
(1 arcsec = 213 pc), this galaxy is physically
very large (diameter, $D\,=\,81$ kpc).  However, its properties
related to star formation (SF) are similar to those of NGC~5907
(see Table~\ref{basic_parameters}) which is half its size.
The SFRs of both
galaxies are modest, i.e. 1 to 3 M$_\odot$ yr$^{-1}$;
 by comparison, the SFR of the better-known
quiescent galaxy, NGC~891, is 3.8 M$_\odot$ yr$^{-1}$ 
(Popescu et al. 2000).  
With an inclination of $i\,=\,90^\circ$ and an SFR that is
similar to a galaxy in which PAH halo emission has already
been detected, 
NGC~5529
is an ideal candidate for the study of extraplanar
emission in general, and the search for high latitude
 PAHs in particular.

To our knowledge, only one recent search for
extraplanar emission in NGC~5529
has been carried out.  This involved an H$\alpha$/N[II] observation
(hereafter referred to as H$\,\alpha$)
by Miller \& Veilleux (2003), with a positive detection.
  These authors report widespread eDIG
within the central 10 kpc of the disk and find a weak, extended
component with an average 
exponential scale height
of $z_e$ = 4.5 kpc.  The total eDIG mass,
M$_{eDIG}$, was found to be
 M$_{eDIG}\,=\,2.0\times\,10^8$ 
M$_\odot$.  By contrast, an  
 early radio continuum observation did not detect extraplanar radio emission,
but the observations were likely not sensitive
enough to do so
(Hummel et al. 1991).  In the soft X-ray band,
an upper limit of $L_X\,=\,2.2\times\,10^{41}$ erg s$^{-1}$ 
(scaled to our adopted distance, and so throughout) has
been placed on the galaxy's total X-ray emission
(Benson et al. 2000). The stellar component of NGC~5529 has been
widely studied, however, and
 various authors have reported a thick stellar disk, although
optical measurements have been hindered, to some extent, by
the prominent dust lane (Fig.~\ref{n5529_colour}).  For example,
 vertical fits to
B, V, and I-band data give 
$z_{s}$ = 0.99, 0.92, and 0.86 kpc, respectively, where the best fit
is to a $sech(|z|/z_s)$ function (de Grijs \& van der Kruit 1996).
More recently, exponential fits to R-band data result in
a mean value of $z_e\,=\,8^{\prime\prime}$ (1.7 kpc), increasing
 with galactocentric distance
 (Schwarzkopf \& Dettmar 2001), an effect that may be
related to the fact that this galaxy is undergoing an interaction
(see below).   
The Two Micron All Sky Survey (2MASS) Ks-band scale height is
$z_{s2}$ = 0.98 kpc (scaled to our distance) where, in this case, 
the luminosity volume is fit by a
$sech^2(|z|/z_{s2})$ function (Bizyaev \& Mitronova 2002).  

In a series of recent papers by Kregel et al. (Kregel et al. 2004a, 2004b, 
Kregel \& van der Kruit 2004, Kregel \& van der Kruit 2005,
and Kregel et al. 2005), which include optical long-slit
spectroscopy and HI observations (HI data are also available from
Rhee \& van Albada 1996), the authors 
 have looked in detail at the structure
and kinematics of edge-on disks, including NGC~5529.
Among their findings are the following.  The NW side of the galaxy is
advancing and the SE is receding with respect to its center.  The
optical distribution is asymmetric, being more extended on the NW
than the SE side (not obvious in Fig.~\ref{n5529_colour} but see
Kregel et al. 2004a). 
The HI distribution shows the classical double-horned profile of 
a rotating disk galaxy, but non-circular motions are seen in the inner parts. 
Of particular
interest is their discovery of HI bridges that connect
 two of the galaxy's companions,
MCG+06-31-085a and Kregel B (both labelled in Fig.~\ref{n5529_colour}),
 to the disk of NGC~5529. Thus, the galaxy
is clearly interacting.  A high velocity dispersion in the disk is further
evidence that the disk is perturbed.

The richness of the field within which NGC~5529 is located may have
some bearing on the activity in NGC~5529 and its halo emission (see
Sect.~\ref{halo_emission}). Therefore, we have listed the group
members and known associated galaxies, from literature sources, in
Table~\ref{group_parameters}.  The total number of previously
known group members is 15 galaxies, of which NGC~5529 is dominant.
In addition, since
NGC~5529 can be found in the Sloan Digital Sky Survey 5th data release
(SDSS DR5, Adelman-McCarthy et al. 2007), we have
 searched this data set for additional group members within 
a square region of angular size,
$27^\prime$ (345 kpc) on a side  
and within a velocity range of $\pm\,500$ km s$^{-1}$ with respect
to NGC~5529.  There are a total of
301 non-stellar objects in this field
(including NGC~5529) of which
35 have spectroscopic redshifts ($z_{sp}$) and all have 
photometric redshifts ($z_{ph}$).  
Although some of the objects which have only photometric redshifts could
be companions, we have discounted these from our search because 
$z_{ph}$ values are unreliable.
For example, the $z_{ph}$ values for NGC~5529
and Kregel B imply incorrect
velocities (see values in parentheses in 
Table~\ref{group_parameters})\footnote{A  
plot of $z_{ph}$ against $z_{sp}$ for the 35 galaxies in the field for which
 both values exist, show typical discrepancies of
$\Delta\,z\,\approx\,0.07$ which is in good agreement with that
found by Fern{\'a}ndez-Soto et al. (2001) for low redshift objects
 in the Hubble Deep Field.}.
Two new companions have been found and are listed as
IKPM~1 and IKPM~2 (along with their SDSS identifiers)
in Table~\ref{group_parameters}.  The newly identified companions are also 
labelled in Fig.~\ref{n5529_colour}.

\section{Observations \& Data Reduction}
\label{observations_reductions}

The ISO $\lambda\,6.75~\mu$m observations were taken on three different days
 (see Table~\ref{observing_map}) using the infrared camera, ISOCAM,
and each data set had a different on-source pointing position.
The observations were taken in 
beam-switching mode which involves obtaining
a sequence of frames on-source, each of 10 s integration time, followed
by a sequence of off-source sky pointings at various positions around the
galaxy.  There were 26 different
 off-source positions, which are listed in Table~\ref{pointings}.  
Each data set contained a total of 675 frames, of which
approximately 40\% were on-source pointings.
  The off-source sky frames were used 
 for the flat field, its distortions, and the mean sky
value at each pixel. This mode of operation provides an accurate means of
making sky and instrumental corrections that are appropriate for
the observing conditions, rather than relying on
a flat-field library, resulting in high dynamic-range
results. 
  Each data set was then reduced separately,
as described below.

All data were reduced using the CAM Interactive Reduction Package
(CIR, Chanial 2003).  First, the dark current was subtracted following
the method of Biviano et al. (1998) which includes a dark correction,
a second-order dark correction depending on detector temperature and
time of observation, and a short-drift correction.  Next, high glitches
due to cosmic ray (CR) hits were removed automatically via a multi-resolution
filtering technique (Starck et al. 1999) at a 6$\sigma$ level.  The effects
of memory on a pixel (transient effects) 
were then corrected (de-glitched) using the 
Fouks-Schubert method (Coulais \& Abergel 2000).  
 Each data set
was then examined carefully, frame by frame, and further bad pixels 
were removed 
manually as required.  Most of these occurred in regions immediately adjacent
to pixels which had been automatically de-glitched.  
An approximate criterion for deglitching was if the pixel value exceeded
$3\sigma$ of the average of the 5 frames that surrounded it in time.
In addition,
although the main effects of memory had been removed as indicated above,
the detector still retained a weak memory of 
the galaxy emission in the first frames of any given off-source
sky field.
Therefore, the first $\approx$ 15 frames
from sky fields were also removed, as were distinct bright objects in
the sky fields.

The on-source frames were then averaged and
corrected for the flat field, its distortions and
sky brightness using the off-source sky frames.  The result
was calibrated to mJy/pixel units following Blommaert et al. (2003).
Formal errors were also carried through the data reduction
on a pixel-by-pixel basis, following the
description in Irwin \& Madden (2006).
 The parameters
of the individual maps and their corresponding error maps are provided
in Table~\ref{observing_map}.  

Each image and its corresponding
error map was then read into the Astronomical Image Processing System
(AIPS) and subsequent analysis carried out in this package. 
The 3 images were compared for consistency in position and flux.
At this point,
the second data set required a slight positional adjustment
 (2.5 pixels in RA) to bring it into alignment with
the other two images. 
We also interpolated over a known bad pixel column.
The three images were then averaged
to form a final map, retaining the original 6$^{\prime\prime}$ pixel
size, and 
the same procedure was then followed for the error maps.
The
parameters of the averaged map and averaged error map
are also given in Table~\ref{observing_map}.
 Note that $\overline\sigma$ represents a typical
formal error in any given pixel, whereas the rms, which is 
lower, is measured directly from the image and represents
 pixel-to-pixel variations in noise 
across the map.
The latter value has been significantly reduced by combining the
three data sets into one.  

The averaged map and its error map
 were then interpolated onto a 1$^{\prime\prime}$ square
pixel grid and rotated to the correct orientation on the sky.
The resulting ISO and 
error maps are shown in Fig.~\ref{iso_map}.
A final 
positional shift of less than a pixel of the original
data (3.4$^{\prime\prime}$ in RA and
5.1$^{\prime\prime}$ in DEC) was then applied to these maps 
to align the ISO map with the SDSS r-band image.  The fine adjustment
in astrometry was possible
because of two sources, shown with crosses
in  the overlay of Fig.~\ref{iso_optical}, that are visible in both
the ISO and optical images.  
Characteristic map errors are also listed in Table~\ref{observing_map}.
In addition, there is an absolute calibration error of order,
$\pm\,15\,-\,20$\% (e.g. Coia et al. 2005,
Pagani et al. 2003), that will affect comparisons with images
at other wavelengths, but should not affect point-to-point comparisons
on the ISO map itself.

\section{Results}
\label{results}

Fig.~\ref{iso_map} shows the final ISO $\lambda\,6.75$ $\mu$m emission
along with its error map, and an overlay of the emission on the SDSS
r-band image is shown in Fig.~\ref{iso_optical}. 
Before studying these data
in detail, however, it is important to establish whether there might
be other significant contributors to the 
$\lambda\,6.75$ $\mu$m emission, besides PAHs. These might include
 an underlying dust 
continuum\footnote{In this paper, we distinguish
the PAH carriers as those components that form spectral
features and refer to an underlying continuum as a possible contribution
from dust whose modified black body
emission increases with increasing wavelength at 
$\lambda\,6.75$ $\mu$m (e.g. Fig.~7, inset, of Irwin \& Madden, 2006).  Such
a continuum may be due to very small grains (VSGs,
Cesarsky et al. 2000).}, 
and a stellar continuum from cool stars.

\subsection{The $\lambda\,6.75~\mu$m Emission}
\label{band_contributions}

Previous observations and spectral modeling indicate that
the MIR spectrum of spiral galaxies is overwhelmingly dominated by
PAH emission 
(e.g. Smith et al. 2007, Draine \& Li 2007,
Vogler et al. 2005, Lu et al. 2003,
Genzel \& Cesarksky 2000 and references therein)
and, as indicated in Sect.~\ref{introduction}, the $\lambda\,6.75$ $\mu$m
ISO band is no exception.  Within this band fall  the 
 $\lambda\,6.2$ $\mu$m,
 $\lambda\,7.7$ $\mu$m, and the short wavelength wing of the
$\lambda\,8.6$ $\mu$m PAH emission features. 
 Of these, the 
$\lambda\,7.7$ $\mu$m feature is the strongest 
(see Fig.~7 of Irwin \& Madden 2006)
and can be modeled as a blend
of three spectral features at
$\lambda\,7.42$ $\mu$m, $\lambda\,7.60$ $\mu$m, and $\lambda\,7.85$ $\mu$m 
(Smith et al. 2007).
 The $\lambda\,7.7~\mu$m PAH complex, alone, can
contribute nearly one-half of the total PAH luminosity and up to 10\%
of the total infrared luminosity (Smith et al. 2007). 
Indeed, photometry of the $\lambda\,6.75~\mu$m ISO band
is normally taken as a direct proxy for the presence of PAH emission.

It has also become clear that galaxies
show qualitatively little difference
in their MIR spectra and that spectral shape is largely independent of
star formation rate 
(Vogler et al. 2005, Lu et al. 2003, Genzel \& Cesarsky 2000
and references therein). Lu et al. (2003), for example, found
band-to-band variations of, at most, 15\% in the mid-IR spectrum
out to $\lambda\,11~\mu$m. 
It is now apparent, however, that some variations in PAH line
ratios can exist, depending
on their environment.  The ratios between the $\lambda\,6.2$,
 $\lambda\,7.7$, and  $\lambda\,8.6~\mu$m PAHs do not seem to show
much variation within galaxies and between galaxies
(Galliano et al. 2007), but the $\lambda\,6.2/\lambda\,11.3$ and
 $\lambda\,7.7/\lambda\,11.3$ ratios can vary by up to an order of
magnitude (Lu et al. 2003; Vogler et al. 2005; Draine \& Li 2007;
Smith et al. 2007; Galliano et al. 2007).  This variation can be
controlled by the fraction of ionized PAHs, which has been demonstrated
to be linked to $G_0/n_e$ (the ratio of the intensity of the UV radiation
field to the electron density, Galliano et al. 2007) or to the
modification of the grain size distribution (Smith et al. 2007).  
For this study, since we are dealing only with the $\lambda\,6.75~\mu$m ISOCAM
band, we can assume constant band ratios between the
$\lambda\,6.2$, $\lambda\,7.7$ and $\lambda\,8.6$ $\mu$m PAH features,
although some variation will not affect our conclusions.

 As for an underlying hot dust continuum, 
such a contribution has been shown to be negligible in normal
star forming galaxies (e.g. Galliano et al. 2007).  For example, 
the three PAH features at 
$\lambda\,6.2$, $\lambda\,7.7$, and $\lambda\,8.6$ $\mu$m 
in M~82 can be fit without any
significant underlying continuum at all
 (Laurent et al. 2000).  Vogler et al. (2005) also
find that a continuum contributes
only $\sim$ 5\% for M~83 in the ISO bands that trace PAH emission.
Depending on the assumption of PAH line profile, recent models
of MIR spectral energy distributions
(SEDs, e.g. Draine \& Li 2007, Galliano et al. 2007) 
typically show an underlying continuum that is
at least an order of magnitude lower than the PAH spectral features. 

A stellar component, on the other hand, may be a somewhat larger
contributor to the observed emission.  Spitzer Space Telescope observations,
for example, are typically corrected for a stellar component
based on the $\lambda\,3.6~\mu$m image which is assumed to contain
only stellar emission.  The correction is based on an extrapolation
of the stellar flux to the relevant wavebands according to the
prescription of Helou et al. (2004) who have used 
 stellar population models and
model photospheric SEDs derived from Starburst99 (Leitherer et al. 1999).
For NGC~5529, the 2MASS Ks band ($\lambda\,2.16~\mu$m) image is available
and we can also assume that this band contains only stellar emission.
Using the same relation of Helou et al., we estimate that the global
flux of NGC~5529 in the $\lambda \,6.75~\mu$m band contains a contribution
of stellar flux that amounts to 14\% of the 
2MASS $\lambda\,2.16~\mu$m flux\footnote{Note that 
 Irwin \& Madden (2006) incorrectly
estimated the stellar flux in their observing bands
for NGC~5907.  A recalculation, based on
 the Helou et al.
(2004) relation, however, results in corrections of
order a few per cent, so their numerical
results and conclusions are not changed.}.


The 2MASS Ks band image, smoothed to the resolution of the
ISO $\lambda \,6.75~\mu$m band and multiplied by 14\% is shown in the
Inset to Fig.~\ref{iso_optical}.  An integration of this map over the
same region as is visible in the $\lambda\,6.75~\mu$m image indicates
that, globally, stellar emission contributes 17\% of the flux in
the ISO $\lambda \,6.75~\mu$m band.  
This is within the range estimated for the absolute calibration error on
typical ISOCAM observations ($\pm\,15\,\rightarrow\,20$\%, 
Sect.~\ref{observations_reductions}) and therefore
is justifiably neglected for the galaxy as a whole. 
In the region of the halo itself,
no stellar halo is observed (Fig.~\ref{iso_optical} Inset)
to the limits of the available 2MASS data.
The rms noise of this map places an
upper limit of $\sigma_*\,=\,0.0002$ mJy arcsec$^{-2}$
on the stellar contribution to the $\lambda \,6.75~\mu$m band, i.e.
an upper limit of approximately 30\% on the lowest contour shown in
Fig.~\ref{iso_map}a, Fig.~\ref{iso_optical} (main) and Fig.~\ref{iso_halpha}
and a lower percentage for the higher contours.
In a rectangular region aligned with the major
axis centered on the north-east halo\footnote{The rectangle was set 
so that it is parallel
to the major axis with a lower edge 8 arcsec (1.7 kpc) above the plane 
in order to
start a full beam width from midplane and be outside of the modeled
disk (see Sect.~\ref{halo_emission}).  The rectangle's size is 121 arcsec 
(25.8 kpc) parallel to the major axis and 61 arcsec (13.0 kpc) parallel
to the minor axis.}
the maximum contribution of the
stellar halo is 7\%.
Since the fraction of stellar emission could vary from location to
location, a possible stellar contribution to the halo is discussed
further in Sect.~\ref{halo_emission}.


\subsection{Disk Emission}

Fig.~\ref{iso_map}a shows strong PAH emission along the
disk of NGC~5529, although the entire disk has not been mapped due to
the truncated field of view.   The emission reaches a maximum at the nucleus
 and a two-dimensional Gaussian fit to this maximum gives
a position of RA = 14$^{\rm h}$ 15$^{\rm m}$ 34.1$^{\rm s}$ $\pm$  0.2$^{\rm s}$,
DEC =  36$^\circ$ 13$^\prime$ 36$^{\prime\prime}$ $\pm$ 2$^{\prime\prime}$,
which agrees with the optical center (Table~\ref{basic_parameters}).
Several other peaks are also observed along the major axis, a double peak
$\approx$ 40$^{\prime\prime}$ from the nucleus along
the north-west major axis and a single
peak $\approx$ 50$^{\prime\prime}$ to the south-east.  This structure is shown
more readily in the plot of intensity
along the major axis shown in Fig.~\ref{slices_fig}a.  Note that there is
no change to the positions and structure described here if a stellar
contribution, as outlined in Sect.~\ref{band_contributions},
 is subtracted from the map,
pixel by pixel.

Irwin \& Madden
(2006) found that the PAH emission in NGC~5907 followed primarily 
the molecular gas distribution in that galaxy, with a fainter wing of emission
extending farther out into the HI dominated region.
To our knowledge, there is no published CO map of NGC~5529 with which
to compare the PAH distribution.
The HI distribution (Kregel et al. 2004b) has
a hole at the nucleus and the HI
emission rises on either side of the nucleus to maxima that are at positions
 farther out
in radius than the peaks of Fig.~\ref{slices_fig}a.  
Therefore, the HI distribution does
not resemble that of the PAHs, to the limits of these measurements.
We
suspect that the bulk of the emission will follow that of the CO,
as in NGC~5907,
once such measurements are made.

\subsection{Halo Emission}
\label{halo_emission}

Fig.~\ref{iso_map}a and especially Fig.~\ref{iso_optical} show that
the PAH emission extends far from the plane of NGC~5529.  This detection
has been possible because of the high sensitivity and dynamic range
(Table~\ref{observing_map}) of the beam-switching observations.
The high-latitude emission shows considerable structure, a result that 
was also seen for NGC~5907. 
Indeed, among other features, 
two discrete arc-like structures can be seen
on the north-east side of the major axis:
 one above the nucleus extending to 
RA $\approx$ 14$^{\rm h}$ 15$^{\rm m}$ 36$^{\rm s}$, 
DEC $\approx$ 36$^\circ$ 14$^\prime$ 10$^{\prime\prime}$, and one
farther to the east extending to 
RA $\approx$ 14$^{\rm h}$ 15$^{\rm m}$ 39$^{\rm s}$, 
DEC $\approx$ 36$^\circ$ 13$^\prime$ 50$^{\prime\prime}$.  Arc-like
features similar to these
have been observed in other ISM tracers in other
edge-on galaxies (e.g. Lee et al.
2001).
At the $3\,\sigma$ contour shown in
Fig.~\ref{iso_optical}, the emission
extends to $z\,\approx\,40^{\prime\prime}$ ($z\,\approx\,8.5$ kpc) above the
plane.  Note that there is no emission (Fig.~\ref{iso_optical} Inset),
to the 2$\sigma$ limit of the {\it stellar map}, to this height.  The maximum
extent of the stellar emission at the 2$\sigma$ limit of this map is 
$z\,\approx\,20^{\prime\prime}$ (4.2 kpc).

Even higher sensitivity can be achieved by
averaging the emission over a broad swath of the major axis
($91^{\prime\prime}$ in width)
and plotting the result as a function of latitude, $z$, above and below
the plane (note that no correction for inclination is required). 
Not only does this approach improve the sensitivity, but it
also allows us to probe whether a broad scale halo exists
around the galaxy, rather than only discrete extensions.
The result is shown in Fig.~\ref{slices_fig}b (main 
figure). Note that the dynamic range of this slice is very high, i.e.
1773/1 (peak/rms).  
Various functions were fitted to these data including
an exponential, a Gaussian, a hyperbolic secant, and 
the square of a hyperbolic secant [that is, so that the range of
functions used by others for optical data
(Sect.~\ref{ngc5529}) are also considered here].
 The best fit was to a simple
Gaussian, the result shown as the grey curve in Fig.~\ref{slices_fig}b
(main).
The fit is excellent, with the exception of the broad scale
wings that are seen in the data but not in the model (see below).
The displayed best fit Gaussian curve,
which is wider than the true distribution because of
beam smoothing, has $\sigma_G\,=\,4.6^{\prime\prime}$
(979 pc), or a full width at half maximum of
FWHM = $10.8^{\prime\prime}$ (2.3 kpc).  We reproduced this curve by 
modeling the $z$ distribution of the disk as a Gaussian of 
FWHM =  $8.0^{\prime\prime}$ (1.7 kpc)
and then convolving with the Gaussian PSF of the observations
(FWHM$_{PSF}$ = $7.2^{\prime\prime}$, Table~\ref{observing_map}).
Thus, after beam correction, the
 bulk of the PAH emission can be represented by a disk 
that declines as a Gaussian of FWHM =  $8.0^{\prime\prime}$ (1.7 kpc),
or $\sigma_G\,=\,3.4^{\prime\prime}$ (726 pc).
We repeated the same exercise over the same region
using the H$\alpha$ data\footnote{The H$\alpha$ data were kindly
made available by Scott Miller.  Note that Miller \& Vielleux (2003) fit
the main disk emission with an exponential; we have repeated our analysis
on their data so as to directly compare best-fit Gaussians.},
 finding
a Gaussian of FWHM =  $13.2^{\prime\prime}$ (2.8 kpc),
or $\sigma_G\,=\,5.6^{\prime\prime}$ (1.2 kpc).  Thus, the H$\alpha$
disk is $\approx\,1.6\,\times$ 
thicker than the PAH disk, when the same regions and criteria
are used.


The faint wings of PAH emission in the very high sensitivity plot of
Fig.~\ref{slices_fig}b (main) reveal
that, as suggested by the appearance of 
Figs.~\ref{iso_map}a and Fig.~\ref{iso_optical},
 a broad scale halo of PAH emission exists around NGC~5529.
The Inset to Fig.~\ref{slices_fig}b
shows the same averaged data from the main figure, but over the north-east
side of the disk only, to avoid contamination by the small galaxy
(likely a background object with $z_{ph}\,=\,0.123$) on the
south-west side.
To the 3$\sigma$ limit of the averaged minor axis slice
(shown by the short horizontal bar),
emission is seen out to  
$z\,\approx\,60^{\prime\prime}$ ($z\,\approx\,12.8$ kpc). 
After subtracting the modeled Gaussian of the prominent disk emission,
the {\it residual} wing emission can be fit with an exponential 
of $z_e\,\approx\,17.5^{\prime\prime}$ (3.7 kpc).  
Miller \& Veilleux (2003) found an exponential fit of
$z_e\,=\,4.5$ kpc
for the high latitude wings of the H$\alpha$ emission, i.e. 
the scale height of the high latitude H$\alpha$ emission is   
$\approx\,1.2\,\times$ larger than that of the PAH emission.

Could the extended $\lambda\,6.7~\mu$m emission be explained 
in some other way than a PAH halo?  As indicated in 
Sect.~\ref{band_contributions}, the global contribution of stellar emission
is within the absolute calibration 
error of the $\lambda\,6.7~\mu$m band emission, if standard extrapolations
are reliable. To further compare the halo stellar emission with that of
the $\lambda\,6.7~\mu$m band, we repeat the same averaging technique
to the Ks band image of Fig.~\ref{iso_optical} (Inset) and we show the
resulting stellar emission profile in Fig.~\ref{slices_fig}b
(Inset, grey curve).  No emission above the 
 3$\sigma$ level of this plot is seen beyond $z\,\approx\,20^{\prime\prime}$
(short horizontal grey line) in contrast to $z\,\approx\,60^{\prime\prime}$
seen at $\lambda\,6.7~\mu$m.  Similar results are obtained when the
SDSS i and z-band images are examined in the same way (slices not shown).
Even if we allow the Ks band emission to contribute to the wings at
approximately a $1\sigma$ level over the entire wing extent, stellar
emission could not account for the $\lambda\,6.7~\mu$m halo.  However,
taking this conservative approach, the maximum extent of the halo
then adjusts to $z\,\approx\,50^{\prime\prime}$ ($z\,\approx\,10.6$ kpc).
Finally, we note that the 
 PSF is known to be Gaussian 
to high accuracy (see Galliano 2004 and
Irwin \& Madden 2006) in this ISO band and the extended PAH
 emission cannot be explained
by PSF emission wings. 


In summary, a significant halo of PAH emission is seen around NGC~5529
and shows considerable substructure with
some features that are vertical or arc-like with respect to the disk.
 The bulk of the emission in the vertical $z$ direction can be fit
by a gaussian with
dispersion, $\sigma_G\,=\,3.4^{\prime\prime}$ (726 pc)
and, after subtracting this main emission, faint PAH wings are seen
with an exponential vertical scale height of 
$z_e\,=\,17.5^{\prime\prime}$ (3.7 kpc).  To the $3\sigma$ limit of
the data and allowing for a small contribution from stars,
emission is seen as far out as $z\,=\,50^{\prime\prime}$
(10.6 kpc).
This is an exceptional
distance from the plane, exceeding that of 
$z\,\approx\,6.5$ kpc found for NGC~5907 (Irwin \& Madden 2006).

\subsection{PAH - H$\,\alpha$ Comparison}
\label{halpha_correlation}

As indicated in Sect.~\ref{ngc5529}, the only previously-observed gaseous halo 
in NGC~5529 was detected 
by Miller \& Veilleux (2003) in H$\,\alpha$ emission,
and we have shown in Sect.~\ref{halo_emission} that 
the vertical distributions in both these bands
can be fit with two vertical components, a narrower Gaussian containing
most of the emission, and fainter exponential wings that extend much farther.
The H$\alpha$ scale heights are, on average, $\approx\,1.4$ times
larger than those of the PAHs when global halo emission is considered.
In this section, we wish to investigate a possible spatial correlation    
between the H$\alpha$ and PAH band emission.  This is shown in
 the two overlays of
 Fig.~\ref{iso_halpha}.

In Fig.~\ref{iso_halpha}a, the H$\,\alpha$ emission is shown in greyscale
with in-disk emission emphasized in order to discern whether the high-latitude
PAH emission may be related to underlying in-disk emission.  
The comparison is not straight-forward since the observed halo structure
results from an integration of emission along lines-of-sight that vary with
radius, the H$\,\alpha$ emission in the disk suffers from extinction, and
the spatial resolutions are different.  The arc
(see Sect.~\ref{halo_emission}) that, in projection, is
located above the nucleus is possibly related to enhanced SF activity in
the nuclear vicinity, but since the extended halo emission is so pervasive,
one-to-one correlations with in-disk activity cannot be pinpointed with
certainty
from these observations.

Fig.~\ref{iso_halpha}b shows the H$\,\alpha$ emission
in contours, smoothed
to the same resolution as the ISO data, over the ISO emission in greyscale.
With the H$\,\alpha$ emission smoothed, the H$\,\alpha$ halo is very
conspicuous and the 'filamentary eDIG' on the northeast side of the galaxy
 noted by
Miller \& Veilleux (2003) is now seen with prominent structure that bears
a remarkable resemblance to that of the PAH emission.  The two PAH halo
arcs are also seen in H$\,\alpha$ as is an above-disk feature located at 
RA $\approx$ 14$^{\rm h}$ 15$^{\rm m}$ 32$^{\rm s}$, 
DEC $\approx$ 36$^\circ$ 14$^\prime$ 40$^{\prime\prime}$.
Similarities on the south-west side of the disk are also apparent.
Thus, although the vertical scale heights of the H$\alpha$ emission
exceed those of the PAHs (Sect.~\ref{halo_emission}), 
there appears to be a spatial correlation between 
the PAH halo structure and that of the 
H$\,\alpha$-emitting eDIG in NGC~5529.  To properly quantify such a correlation
requires a three-dimensional model of the two components and the
low S/N of the data in the halo region does not support such an approach.  
However, to `zeroth order', we have investigated whether the two maps
are correlated  in the region of the  
the north-east halo. After applying a $2\,\sigma$ cutoff to both maps, the
resulting halo emission spans a projected area of 0.79 square arcminutes.
For this region, we computed cross-correlation coefficients between
the two maps, finding a best value of
 95\% for zero shift in position.  This confirms
 that there is a correlation between the PAH and H$\alpha$ halo emission
in NGC~5529.

\section{Discussion}
\label{discussion}

The presence of a large-scale, structured PAH halo about NGC~5529 
is a significant result of these observations.
Few statistics exist on the presence of PAHs in galaxy halos.
Of normal star forming galaxies or those of low SFR, NGC~5529 and
NGC~5907 appear to be the only known examples, thus far
(see Table~\ref{basic_parameters}).
Tacconi-Garman et al. (2005) have found a $\lambda\,3.3~\mu$m PAH feature
in the superwind of 
the starburst galaxy, NGC~253, with $z$ extent $< 120$ pc, and there is now a
clear PAH signature in the halo and
superwind of M~82 to
a distance of 6 kpc from the plane of that galaxy (Engelbracht et al. 2006).  
Its 8 $\mu$m
emission, which has a strong PAH component, resembles the H$\alpha$ emission
of the superwind in M~82, 
indicating that PAHs can survive in such an energetic environment
and providing another example of a galaxy in which PAHs correlate with
H$\alpha$ emission in the halo.
Engelbracht et al. also find that filamentary 8 $\mu$m 
emission is present, not only in the superwind outflow, but also throughout
the halo region, as we see for NGC~5529.  They attribute the widespread
halo PAHs 
to previous activity related to star formation.
 Thus
halo PAHs are now seen in galaxies with both high and
low SFRs.

PAHs are believed to originate in the dusty envelopes of evolved AGB stars 
(e.g. Boersma et al. 2006, Matsuura et al. 2004, Cherchneff 1995) 
and there is now also good evidence that PAHs may be formed from
the destruction of small grains 
(Cesarsky et al. 2000, Rapacioli et al. 2005, Bern{\'e} et al. 2007).
Indeed, PAHs which have fewer than
$5.75\,\times\,10^4$ carbon atoms have radii 
less than 0.005 $\mu$m (Draine \& Li 2007) and are therefore
similar in size to  the
very small grains (VSGs) that were first introduced to explain 
the 12 $\mu$m diffuse emission seen in the Milky Way.

In either case, PAHs are expected to reside in galaxy disks, rather than
their halos, unless a distribution of halo AGB stars were the origin of
the PAHs. We have insufficient theoretical and observational
data to quantify
the latter possibility\footnote{To quantify this requires 
  a knowledge of the halo AGB
star population, 
a knowledge of the PAH formation rate 
(metallicity-dependent) and destruction rate,
a SED model that predicts the strength of the various PAH features taking
into account stochastic single-photon excitation from in-disk photons 
(dependent on vertical optical depth), 
 and some indication as to how these quantities
vary with time.
  See Temi et al. (2007a, 2007b)
for a similar approach, but for the different 
conditions in elliptical galaxies, for
classical grains rather than PAHs, and assuming dominant destruction by
spallation in a hot X-ray environment.} 
but we note that
 the presence of structured, vertical and arc-like
features connected to the disk in both NGC~5907 and NGC~5529
 argue that halo PAHs originate
in the underlying galaxy disk and are ejected into the halo.
We expect, for example, that an origin in halo AGB stars would result 
in a more uniform distribution of halo emission.  
Furthermore, for NGC~5529 in particular, the PAH - H$\,\alpha$
correlation (Sect.~\ref{halpha_correlation} and below) argues for a disk 
origin since it is widely believed, based on correlations with
disk SFRs for example, that the eDIG in galaxies 
originates in the in-disk ISM.
We therefore assume that the
halo PAHs of NGC~5529 originate in its disk, in agreement with 
 Engelbracht et al. who similarly argue for a disk origin for
the halo PAHs of M~82 that are unrelated to the
nuclear outflow.


In the Milky Way, we know that PAH emission tends to be strongest in
photodissociation regions (PDRs) since, in such regions, there are
sufficient far-UV (FUV) photons to excite the PAHs
 but not so many that
the probability of photo-destruction becomes high (see below).
 For example, the Orion bar PDR shows a strong PAH signature
(e.g. Sloan et al. 1997) and 
shells of PAH emission can be seen around HII
regions in the star forming complex, NGC~6334 
(Burton et al. 2000).  In HII regions, themselves, PAH
emission is suppressed because PAH molecules cannot
survive for long in such a harsh environment (e.g. Dopita et al. 2005).
On the other hand, some PAH emission has indeed been observed in
ionized regions such as planetary nebulae
and HII regions (e.g. Peeters et al. 2005, van Diedenhoven et al. 2004,
Vijh et al. 2004).

The survival  and emission properties of PAHs in various interstellar 
environments has now been modeled by 
a number of authors (Allain et al. 1996a, 1996b,
Le Page et al. 2001, 2003, Li \& Draine 2001,
Draine \& Li 2007).  
The most relevant process, when considering the survival of PAHs
in galaxy halos, is the photo-destruction rate
(we do not consider the 
possible growth of PAHs through accretion of carbon atoms in the halo
environment).
Other destruction processes, such as 
collisional destruction in shocks or sputtering from the presence of
hot X-ray emitting gas
seem less likely, the former because of
the halo environment, and the latter because
no X-ray halo has yet been observed in this galaxy
 (Sect.~\ref{introduction}).
PAHs can be photo-dissociated by the loss of a hydrogen atom, a carbon
atom, C$_2$H$_2$, or H$_2$.  The type and rate of photo-destruction is
a function of PAH size (i.e. number of carbon atoms, $N_C$)
and the strength and energy of the radiation field.  
 Large
PAHs are more stable than smaller PAHs, in general, and 
the loss of each particular particle
has a different threshold energy for ejection, typical values being
in the FUV range, 2 to 8 eV.

The lifetime of a large PAH against photo-ejection of particle, $i$, can
be estimated from,
\begin{equation}
\tau\,=\, \frac{1}{\chi\,k_{ISM}^{i}}
\end{equation}
(adapted from Allain et al. 1996a),
where $k_{ISM}^{i}$ (s$^{-1}$)
is the rate of PAH destruction by the ejection of particle, $i$,
 in the Milky Way's ISM
 and $\chi$ is a factor that accounts for the difference
in the strength of the radiation field between the ISM 
of the Milky Way and the environment
of interest.  The rate, $k_{ISM}^{i}$, is a function of $N_C$
as well as the threshold photon energy for the ejection of species, $i$.
For $N_C \,=\,50$, for example, $k_{ISM}^{C_2H_2}\,=\,3.56\,\times\,10^{-18}$
s$^{-1}$.  The highest destruction rate 
for $N_C \,=\,50$
can be attributed to the
ejection of H, for which 
$k_{ISM}^{H}\,=\,2.3\,\times\,10^{-16}$
s$^{-1}$ (see Allain et al. 1996a).
 For the halo of NGC~5529, we can set a lower limit to
the lifetime of a 50 carbon atom PAH by assuming that the radiation field
is as strong as the ISM in the Milky Way ($\chi\,=\,1$) and that
destruction is via H ejection.  The resulting lifetime is
 $\tau\,=\,1.4\,\times\,10^8$ yr.  
Thus, even for an unrealistically strong radiation field,
such a PAH would be very long lived.  Assuming a more
realistic halo radiation field in which FUV photons leak from the
disk into the halo
(adopting $\chi\,=\,0.1$ as a possible example) 
then $\tau\,=\,1.4\,\times\,10^9$ yr.
In addition, some potentially photo-dissociating FUV photons 
will instead be absorbed by hydrogen (see, e.g.
Dopita 2003, 2005), leading to a longer lifetime.  Therefore, we
 expect halo PAHs of this size to be stable over long
time periods.  Note, however, that we have
not considered any re-settling of PAHs back onto the disk as
might occur from a galactic fountain.

We can also ask what vertical velocity, $v_z$,
would be required in order to
eject PAHs 
from the disk to the halo of NGC~5529 to a height of $\approx\,10$ kpc 
(Sect.~\ref{halo_emission}) before the PAHs would be photo-dissociated.
In the strong radiation field limit ($\tau\,=\,1.4\,\times\,10^8$ years)
the result is 
$v_z\,\approx\, 70$ km s$^{-1}$.  For the more realistic weaker
radiation field ($\tau\,=\,1.4\,\times\,10^9$ years),
$v_z\,\approx\, 7$ km s$^{-1}$ is sufficient.  The latter value is
too low to actually eject material into the halo, but the main
 conclusion is that PAHs do not need
to be supplied into the halo at high velocities in order to account for
the observed emission at high galactic latitudes.  
Since PAHs are long-lived, it is only necessary that there has been
sufficient energy to eject the particle into the halo, either at some
point in the past or more quiescently over a period of time.
This is consistent with the fact that we see PAHs in 
the halos of galaxies with
low SFRs.  Moreover, if this is correct, then most star forming
galaxies should show some PAH halo emission.
The presence of vertical structures (or vertical filamentary features,
as described for M~82) suggests that the ejection
in NGC~5529 likely did not occur
more than $\approx\,10^8$ years ago; otherwise, galactic rotation
would destroy these structures.
The process by which PAHs are transported to such high galactic latitudes
needs to be more fully addressed by theory and observation. 
We know that stellar winds and supernovae can contribute to vertical
outflows in galaxies
and that the outflows will include entrained ISM material.  Vertical magnetic
fields can also play an important role in more quiescent ejection of
ISM material.  Just how PAHs, either neutral or ionized, might couple
to the outflow in either case is not yet clear.  In the case of NGC~5529,
its interaction
with one or more companions may also have played a role in agitating
the ISM.


Stellar photons over a range of energy are likely leaking from the
disk into the halo regions.  The lower energy optical and FUV
photons will excite the PAH emission, resulting in the observed
MIR bands.  Some FUV photons of sufficient
energy may also result in photo-dissociation, but at a slow rate
(dependent on $N_C$).  Photons with energies  
 that are higher than
13.6 eV can ionize hydrogen, producing the observed 
H$\alpha$ emission.  Although, in HII regions in the Milky Way,
PAHs are readily destroyed, we note that the radiation field is
much stronger in HII regions ($\chi\,\approx\,10^3$) than in the
diffuse ISM ($\chi\,=\,1$) and higher still than in a halo environment
(Allain et al. 1996a).  
Therefore, we would not expect the halo PAHs to avoid H$\alpha$ emission
as they do in the disk ISM.
The observed PAH/H$\alpha$ correlation in the
halo of NGC~5529, then, may simply represent the fact that gas and
dust are correlated in the halo (as they are in the disk) and that the
sources of excitation (i.e. photons from stars in the disk) are the same.

\section{Conclusions}
\label{conclusions}

NGC~5529 is a large (diameter of 81 kpc), 
 edge-on ($i\,=\,90^\circ$) galaxy with a low
star formation rate (SFR $1.7$ M$_\odot$ yr$^{-1}$).  
It resides 
in a small galaxy group and we have now consolidated a list of
previously known companions from a variety of catalogues. In addition,
we identify two new companions, based on their proximity to NGC~5529 and
their spectroscopic redshifts, as listed in the Sloan Digital
Sky Survey (Data Release 5).  In total, there are now at least 17 galaxies in
the group, of which NGC~5529 is the dominant member 
(Table~\ref{group_parameters}).

We have presented ISO beam-switching $\lambda\,6.7~\mu$m observations 
of the edge-on galaxy, NGC~5529.  These data have very high dynamic range
(530/1)
and have allowed us to probe both the disk and faint halo region of
the galaxy.  The $\lambda\,6.7~\mu$m band is dominated by PAH emission
and, globally, it is unlikely that stellar emission constitutes more than
17\% of the total.  

The disk PAH emission does not resemble the HI distribution. 
It peaks at the optical nucleus, whereas the HI distribution has a central
hole. Molecular
data are lacking, but we expect that the in-disk distribution
will follow
the molecular gas component as seen in other galaxies.  
The vertical
disk distribution is well fit by a Gaussian with $\sigma_G\,=\,726$ pc
(FWHM$\,=\,1.7$ kpc) after correction for the beam.  

In addition, very
high latitude PAH emission is also observed, forming a `halo' around the
galaxy with considerable substructure, including vertical features connected to
the disk (Fig.~\ref{iso_optical}), arguing for an in-disk origin.
The halo PAH emission extends to 
$z\,\approx\,10$ kpc, to the limits of the noise on an averaged
vertical slice (dynamic range = 1773/1).  
 This is the most extensive
PAH halo yet observed in a galaxy. The vertical exponential scale
height of the extended halo component is $z_e\,=\,3.7$ kpc
which is simewhat smaller than the value of 
$z_e\,=\,4.5$ kpc found by Miller \& Vielleux (2003)
for the H$\alpha$ emission in this galaxy.
When the H$\alpha$ distribution is
smoothed to the ISO resolution, however, there is a spatial
correlation between the PAH and H$\alpha$ halo emission.

PAHs can be long-lived in a halo environment which implies that,
provided there has been sufficient energy to eject particles into the halo
and there are currently
sufficient photons leaking from the disk to excite them, a high
SFR is likely not necessary for their detection in galaxy halos.
Given the very low excitation conditions in galaxy halos in comparison
to in-disk HII regions, we do not expect that PAHs will be destroyed
in regions in the ionized halo gas as they are in HII regions in the disk.
The PAH - H$\alpha$ correlation may result from the fact that
both PAHs and ionized components are, to some extent, tracing  
 the actual density distribution in the
halo, with
excitation from a similar underlying in-disk stellar component.
The optical and FUV photons that leak into the halo will
excite PAH emission and harder UV
photons will ionize the gas.

\begin{acknowledgements}
We are grateful to Scott Miller for making the H$\,\alpha$ image of
NGC~5529 available to us.  JI would like to thank the Natural Sciences
and Engineering Research Council of Canada for her research grant.
\end{acknowledgements}

\clearpage

\begin{figure*}
\centering
\caption{Optical image of NGC~5529 and its near environment, from 
SDSS DR5. The displayed field of view is $11.95^\prime$ on a side. Companions
to NGC~5529 are marked in white (see Table~\ref{group_parameters}) and background
galaxies in this field for which spectroscopic redshifts are known
are marked in yellow.  The spectroscopic redshift of NGC~5529 is
$z\,=\,0.00959$.}
\label{n5529_colour}
\end{figure*}

\clearpage

\begin{figure*}
\centering
\caption{
(a) Final ISO $\lambda\,6.7$ $\mu$m map of NGC~5529 in both contours
and greyscale.  The greyscale is over a linear range from 
-0.005 to 0.200 mJy arcsec$^{-2}$ and the truncated boundary shows
the region of sky  that was mapped.  Contours are at 0.0006 (2$\sigma$),
0.0009, 0.0014, 0.0023, 0.0040, 0.0100, 0.0200, 0.0500, 0.1000, and 0.1500
mJy arcsec$^{-2}$. Small tick marks indicate a declining contour.
(b)  Error map, as described in the text, shown in
linear greyscale from -0.0001 to 0.008 mJy arcsec$^{-2}$. The straight
white bar stretching from top center to lower left shows the location of
the bad column on the detector.}
\label{iso_map}
\end{figure*}

\clearpage

\begin{figure*}
\centering
\caption{
ISO $\lambda\,6.75$ $\mu$m image, shown with the same contours as
in Fig.~\ref{iso_map}a, overlaid on the SDSS r-band iamge.
Fine adjustments to the astrometry were made using the objects marked
with crosses, i.e.  
the star at the top of the field and the galaxy just below NGC~5529
identified as 
J141534.3+361304.5 in the SDSS DR5
 (likely a background
source with a photometric redshift of
$z_{ph}\,=\,0.123$). The scale is marked with a vertical bar.
{\bf Inset:}  2MASS Ks band image of NGC~5529,
smoothed to the same resolution as the ISO $\lambda\,6.75$ $\mu$m image,
and multiplied by a factor of 0.14 to represent the approximate
strength of the stellar emission at $\lambda\,6.75~\mu$m (see
Sect.~\ref{band_contributions}).  Contours are the same as for the
ISO $\lambda\,6.75$ $\mu$m image except for the addition of the lowest
contour of 0.0004 mJy arcsec$^{-2}$ (2$\sigma$).}
\label{iso_optical}
\end{figure*}

\clearpage

\begin{figure*}
\centering
\caption{(a) Intensity distribution of
 the major axis of NGC~5529, averaged over a strip centered at
mid-plane and of vertical width, 7$^{\prime\prime}$.
 The center is at the galaxy's nucleus and positive values correspond
to the south-east side of the galaxy. 
(b) Intensity distribution along the
minor axis of NGC~5529, averaged over a strip centered at
the galaxy's center and of width,
91$^{\prime\prime}$, along the major axis.  The dynamic range of this
slice is 1773/1. 
Positive values correspond to the north-east side of the galaxy's disk.
The grey curve is the best fit Gaussian (see text).
(Inset) Blow-up of the minor axis distribution in the direction towards
the north-east only (black curve).
The axes have the same units as the larger figure. The gray curve is
the minor axis slice of the Ks-band image shown in the Inset
to Fig.~\ref{iso_optical} averaged over the same region as the
$\lambda\,6.7~\mu$m image for comparison.  The two short horizontal
bars mark the 3$\sigma$ levels of the $\lambda\,6.7~\mu$m minor
axis slice (black, 1$\sigma$ = 0.073 $\mu$Jy arcsec$^{-2}$)
and Ks-band slice (grey,
 1$\sigma$ = 0.18 $\mu$Jy arcsec$^{-2}$), respectively.  
}
\label{slices_fig}
\end{figure*}

\clearpage

\begin{figure*}
\centering
\caption{(a) ISO $\lambda\,6.7$ $\mu$m image, shown with the same contours as
in Fig.~\ref{iso_map}a, overlaid on a linear greyscale
H$\,\alpha$ image of Miller \& Veilleux
(2003).  The H$\,\alpha$ image is in arbitrary units and is shown to emphasize
detail in the disk.
(b) Here, the contours (arbitrary units) are of H$\,\alpha$ emission,
 smoothed to 7.2$^{\prime\prime}$ resolution to emphasize high latitude emission,
 overlaid on the
 ISO $\lambda\,6.7$ $\mu$m image in linear greyscale (greyscale range =
$-0.003$ to $0.10$ mJy/arcsec$^{-2}$).}
\label{iso_halpha}
\end{figure*}


\clearpage
\begin{table*}
\begin{minipage}[t]{0.57\textwidth}
\caption{Galaxy Parameters}
\label{basic_parameters}
\renewcommand{\footnoterule}{}
\begin{tabular}{lcc}
\hline \hline
Parameter\hfill      & NGC~5529\footnote{Values  
from the NASA Extragalactic Database (NED) unless otherwise indicated.}
			           & NGC~5907\footnote{Values from Irwin \& Madden (2006), or calculated from this table, unless otherwise indicated.}\\
\hline	
RA (J2000) (h m s)   & 14 15 34.07 & 15 15 53.69\\
DEC (J2000) ($^\circ$ $^\prime$ $^{\prime\prime}$) 
		     & 36 13 35.7  &  56 19 43.9\\
V$_{hel}$\footnote{Heliocentric radial velocity.} 
       (km s$^{-1}$) & 2875      &  667\\
Distance (Mpc)       & 43.9\footnote{From Tully (1988).  The distance to the group, GH~141, is
 $D\,=\,49.1$ Mpc (van Driel et al. 2001).  We assume
$H_0\,=\,75$ km s$^{-1}$ Mpc$^{-1}$ throughout.}
				   & 11  \\
Morphological Type   & Sc          & SA(s)c: sp     HII:  \\
$2a$ $\times$ $2b$ \footnote{Major $\times$ minor axis angular size.}
 ($^\prime$ $\times$ $^\prime$)
		     &6.35 $\times$ 0.64 &  12.77 $\times$  1.40 \\
$D$\footnote{Major axis linear size.} (kpc) 	     
		     & 81          & 41  \\
Inclination ($^\circ$) & 90 \footnote{From Tully (1988).}
				   & 86.5 \\
$f_{12}$, $f_{25}$, $f_{60}$, $f_{100}$ \footnote{Infrared fluxes at 
$\lambda\,12$, $25$, $60$ and $100$ $\mu$m, respectively.} (Jy) 
		     & 0.255, 0.234, 1.95, 7.73 
				   & 1.29, 1.44, 9.14, 37.43 
\footnote{From Sanders et al. (2003).} \\ 
$f_{60}/f_{100}$     & 0.25        & 0.24 \\
$L_{FIR}$, $L_{IR}$\footnote{Far infrared (FIR) and infrared (IR) luminosity, respectively, 
using the formalism of Sanders \&
Mirabel (1996) with their constant, $C$ = 1, and 
$L_\odot\,=\,3.84\times\,10^{33}$ erg s$^{-1}$.}
     ($L_\odot$)     & 9.7 $\times$ 10$^9$, 1.9 $\times$ 10$^{10}$
   				   & 2.9 $\times$ 10$^9$, 5.8 $\times$ 10$^{9}$
\footnote{Note that the FIR luminosity differs slightly from the value of 
Irwin \& Madden (2006) since the latter authors use a different
formalism than described in note $j$.} \\
$L_{FIR}/A$ \footnote{FIR luminosity per unit area, where $A\,=\,\pi\,D^2/4$. See also
note $k$.} (erg s$^{-1}$ kpc$^{-2}$)
		     & 7.2 $\times$ 10$^{39}$ 
				   & 8.4 $\times$ 10$^{39}$ \\
SFR \footnote{Star formation rate, using $L_{IR}$ and the formalism of
Kennicutt (1998).} (M$_\odot$ yr$^{-1}$) 
		     & 3.3         & 1.0 
\footnote{This value could be as high
as SFR = 2.2 M$_\odot$ yr$^{-1}$ (Misiriotis et al. 2001).} \\
SFR/A \footnote{See note $l$.} (M$_\odot$ yr$^{-1}$ kpc$^{-2}$)
		     & 6.5$\times\,10^{-4}$ 
				   & 7.6$\times\,10^{-4}$ \\
\hline
\end{tabular}
\end{minipage}
\end{table*}

\clearpage
\begin{table*}
\begin{minipage}[t]{\textwidth}
\caption{Galaxy Group Membership$^1$}
\label{group_parameters}
\centering
\renewcommand{\footnoterule}{}
\begin{tabular}{lccccccc}
\hline \hline
Galaxy Name\footnote{Names beginning with J are identifiers from the SDSS DR5.   Those
beginning with I are our labels.}\hfill 
	& RA (J2000)  & DEC (J2000) 
				 & Velocity\footnote{Heliocentric.  Values in parentheses have been computed from SDSS photometric redshifts.} 
					& Major Axis 
					       & Separation & Group ID\footnote{Group Membership.  GH = Geller \& Huchra,  TG = Turner \& Gott,NOGG = Nearby Optical Galaxies Group member (hierarchical algorithm).} 
									        & Ref.
\footnote{Reference for Group Membership or association with NGC~5529.
1: van Driel et al. (2001), 2: Turner \& Gott (1976),
3: Giuricin et al. (2000)
4: Kregel et al. (2004b)}\\
         & (h m s)    & ($^\circ$ $^\prime$ $^{\prime\prime}$) 
				 & (km s$^{-1}$) 
					& (arcmin) 
					       & (arcmin) &                      & \\
\hline
NGC~5529 & 14 15 34.1 & 36 13 36 & 2875 (32,770) & 6.35 & ---      & GH141, TG86, NOGG773 & 1, 2, 3 \\
NGC~5533 & 14 16 07.7 & 35 20 38 & 3866 & 3.1  & 53.4     & GH141, TG86          & 1, 2 \\
NGC~5544 & 14 17 02.5 & 36 34 18 & 3040 & 1.08 & 27.3     & GH141, TG86, NOGG773 & 1, 2, 3 \\
NGC~5545 & 14 17 05.2 & 36 34 31 & 3079 & 1.3  & 27.8     & GH141, TG86, NOGG773 & 1, 2, 3 \\
NGC~5557 & 14 18 25.7 & 36 29 37 & 3213 & 2.3  & 38.1     & GH141, TG86          & 1, 2 \\
NGC~5589\footnote{Also listed as NGC~5588.}
	 & 14 21 25.1 & 35 16 14 & 3397 & 1.09 & 91.4     & GH141 		 & 1 \\
NGC~5590\footnote{Also listed as NGC~5580.} 
	 & 14 21 38.4 & 35 12 18 & 3221 & 1.8  & 96.1     & GH141, TG86          & 1, 2 \\
NGC~5596 & 14 22 28.7 & 37 07 20 & 3122 & 1.13 & 99.0     & GH141                & 1 \\
NGC~5614 & 14 24 07.6 & 34 51 32 & 3892 & 2.5  & 132.8    & GH141, TG 86 	 & 1, 2 \\
NGC~5656 & 14 30 25.5 & 35 19 15 & 3163 & 1.66 &  188.8   & GH141 		 & 1 \\
NGC~5675 & 14 32 39.8 & 36 18 08 & 3973 & 2.12 & 206.8    & GH141 		 & 1 \\
NGC~5684 & 14 35 50.2 & 36 32 36 & 4082 & 1.56 & 245.5    & GH141 		 & 1 \\
NGC~5695 & 14 37 22.1 & 36 34 04 & 4225 & 1.3  &  264.0   & GH141 		 & 1  \\
MCG~+06-31-085a 
	 & 14 15 19.1 & 36 12 08 & 2968 & 0.47 & 3.3      & 			 & 4\\
Kregel B\footnote{Position and velocity from Ref.~4 (see $e$).}
	 & 14 15 35.5 & 36 12 02 & 2959 (447) 
					&      & 1.6 	  &  			 & 4\\
\hline
\multicolumn{7}{c}{\bf NEW COMPANIONS}\\
\hline
IKPM 1   & 14 15 43.5 & 36 13 19 & 3038 & 0.21 & 1.9 	  & 			 & this work\\
J141543.48+361319.1
	 &	      &		 &	&      &	  &			 &\\
IKPM 2	 & 14 16 00.6 & 36 17 10 & 2645 & 0.50 & 6.4 	  & 			 & this work\\
J141600.64+361710.4
	 &	      &		 &	&      &	  &			 &\\
\hline
\end{tabular}
\leftline{~~~$^1$Data from NED unless otherwise indicated.}
\end{minipage}
\end{table*}

\clearpage
\begin{table*}
\begin{minipage}[t]{0.82\textwidth}
\caption{Observing \& Map Parameters}
\label{observing_map}
\renewcommand{\footnoterule}{}
\begin{tabular}{lcccc}
\hline\hline
Parameter\hfill & TDT 56100405 \footnote{TDT no. is a unique number that identifies the ISO observation.}
				  & TDT 56300315 $^a$ & TDT 56401011 $^a$ & Final Map \\
\hline
Central Wavelength\footnote{Blommaert et al. (2003).} [Range] ($\mu$m) 
		&  6.75 [5.0-8.5] & 6.75 [5.0-8.5]    & 6.75 [5.0-8.5]    & 6.75 [5.0-8.5] \\
Date of Observations 
		& 30/May/1997     & 01/June/1997      & 02/June/1997 	  & \\
Observing Mode$^{b}$\footnote{CAM03 is a beam-switching mode for photometric imaging.}
	        & CAM03 	  & CAM03 	      &CAM03 		  &CAM03 \\
Pixel Field of View \footnote{Size of a square pixel on the sky.} (arcsec) 
		& 6.0 		  & 6.0 	      & 6.0 		  & 6.0 \\
PSF (FWHM)\footnote{Full width at half maximum of the point spread function.} (arcsec) 
		& 7.2 		  & 7.2 	      & 7.2		  & 7.2 \\
Frame Field of View \footnote{Size of the square frame which contains an array of 32 
$\times$ 32 pixels.  The final image size is slightly smaller due to editing.} (arcmin) 
		& 3.2 		  & 3.2 	      & 3.2 		  & 3.2 \\
No. of Frames \footnote{Number of frames in the data set. Approximately 40\% of these frames were on-source.}     & 675 		  & 675 	      & 675 		  & \\
Integration Time per Frame (s) 
		& 10.08 	  & 10.08 	      & 10.08 		  & \\
Calibration Error$^b$ (\%)
		& 3.3 		  & 3.3 	      & 3.3 		  &\\
$\overline\sigma$, rms$_\sigma$ \footnote{Average $\sigma$ and rms in $\sigma$, respectively, over the field, where the $\sigma$ is the formal random error carried through by the software throughout the data reductions. A typical error in a single pixel value would be given by $\overline\sigma$.} (mJy pixel$^{-1}$) 
		& 0.058, 0.083    & 0.035, 0.026      & 0.032, 0.030      & 0.044, 0.039 \\
Peak map flux (mJy pixel$^{-1}$) 
		& 6.50 		  & 7.57 	      & 8.23 		  & 7.43 \\
~~~~~~~~~~~~~~(mJy arcsec$^{-2}$) 
		& 		  & 		      & 		  & 0.203 \\
Rms\footnote{Measured rms noise level of the map.} 
     (mJy pixel$^{-1}$) 
		&  0.020 	  & 0.021 	      & 0.033 		  & 0.014 \\
~~~~~(mJy arcsec$^{-2}$) 
		& 		  & 		      & 		  & 0.0003 \\
Dynamic Range\footnote{Peak map flux divided by rms.}
	        & 325 		  & 360 	      & 249 		  & 531 \\
\hline
\end{tabular}
\end{minipage}
\end{table*}

\clearpage
\begin{table*}
\begin{minipage}[t]{0.4\textwidth}
\caption{Off-source Pointing Positions}
\label{pointings}
\centering
\renewcommand{\footnoterule}{}
\begin{tabular}{cc}
\hline \hline
Right Ascension (J2000)\hfill & Declination (J2000)\hfill \\
 (h m s) & ($^\circ$ $^\prime$ $^{\prime\prime}$)\\
\hline
14 15 36.10 & 36 18 34.8 \\
14 15 32.65 & 36 23 16.8 \\
14 15 35.47 & 36 22 15.9 \\
14 15 34.87 & 36 21 08.3 \\
14 15 31.95 & 36 15 55.6 \\
14 15 35.34 & 36 14 18.3 \\
14 15 35.64 & 36 14 10.5 \\
14 15 32.87 & 36 10 02.9 \\
14 15 33.74 & 36 06 44.9 \\
14 15 36.27 & 36 07 58.0 \\
14 15 33.69 & 36 14 33.9 \\
14 15 34.15 & 36 14 42.3 \\
14 15 36.49 & 36 14 13.0 \\
14 16 00.34 & 36 14 41.6 \\
14 16 12.24 & 36 15 25.2 \\
14 16 15.29 & 36 14 19.5 \\
14 16 04.60 & 36 14 39.8 \\
14 15 34.08 & 36 15 29.0 \\
14 15 34.03 & 36 14 41.1 \\
14 15 34.08 & 36 14 18.0 \\
14 15 04.45 & 36 15 16.0 \\
14 14 51.92 & 36 15 17.1 \\
14 14 59.43 & 36 14 13.9 \\
14 15 26.41 & 36 15 01.3 \\
14 15 32.83 & 36 15 12.4 \\
14 15 35.66 & 36 14 12.9 \\
\hline
\end{tabular}
\end{minipage}
\end{table*}

\end{document}